\documentclass[10pt]{iopart}

\usepackage{mathrsfs}
\usepackage{iopams}
\usepackage{cite}
\usepackage{epsfig}
\usepackage{graphicx}
\usepackage{color}

\def\eps{\epsilon}
\def\Th{\Theta}
\def\sig{\sigma}

\def\3nab{\tilde{\nabla}}

\def\hsp5{\hspace{5mm}}
\newcommand{\sfrac}[2]{{\textstyle{#1\over#2}}}
\def\case#1/#2{\textstyle\frac{#1}{#2}}

\def\be {\begin{equation}}
\def\ee {\end{equation}}
\def\bea {\begin{eqnarray}}
\def\eea {\end{eqnarray}}

\def\case#1/#2{\textstyle\frac{#1}{#2} }
\def\rf#1{(\ref{#1})}

\def\equi {equilibrium }

\def\cqg{{\em Class. Quantum Grav.\/} }

\def\prd{{\em Phys. Rev.\/} {\bf D}}
\def\prl{{\em Phys. Rev. Lett.\/} }
\def\apj{{\em Astrophys. J.\/} }

\def\aph{{\em Ann. Phys. (NY)\/} }
\def\plb{{\em Phys. Lett.\/} {\bf B}}
\def\npb{{\em Nucl. Phys.\/} {\bf B}}

\begin{document}
%%%%%%%%%%%%%%%%%%%%%%%%%%%%%%%%%%%%%%%%%%%%%%%%%%%%%%%%
\title{Compactifying the state space for alternative theories of gravity}
%%%%%%%%%%%%%%%%%%%%%%%%%%%%%%%%%%%%%%%%%%%%%%%%%%%%%%%%%
\author{Naureen Goheer \dag, Jannie A. Leach \dag \ and Peter K.S. Dunsby \dag \ddag}
\address{\dag \ Department of Mathematics and Applied Mathematics, University of Cape Town, Rondebosch,
7701, South Africa}

\address{\ddag \  South African Astronomical Observatory, Observatory, Cape Town, South Africa}

\date{\today}

\eads{ngoheer@yahoo.com, jannie.leach@uct.ac.za and
peter.dunsby@uct.ac.za}
%%%%%%%%%%%%%%%%%%%%%%%%%%%%%%%%%%%%%%%%%%%%%%%%%%%%%%%%%%%
\begin{abstract}
In this paper we address important issues surrounding the choice of
variables when performing a dynamical systems analysis of
alternative theories of gravity. We discuss the advantages and
disadvantages of compactifying the state space, and illustrate this
using two examples. We first show how to define a compact state
space for the class of LRS Bianchi type I models in $R^n$-gravity
and compare to a non--compact expansion--normalised approach. In the
second example we consider the flat Friedmann matter subspace of the
previous example, and compare the compact analysis to studies where
non-compact non--expansion--normalised variables were used. In both
examples we comment on the existence of bouncing or recollapsing
orbits as well as the existence of static models.
\end{abstract}
%%%%%%%%%%%%%%%%%%%%%%%%%%%%%%%%%%%%%%%%%%%%%%%%%%%%%%%%%%
%%%%%%%%%%%%%%%%%%%%%%%%%%%%%%%%%%%%%%%%%%%%%%%

%%%%%%%%%%%%%%%%%%%%%%%
\section{Introduction}
%%%%%%%%%%%%%%%%%%%%%%%

Over the past decade a number of models have been proposed to
account for the shortcomings of the standard model of cosmology
based on General Relativity (GR). In most of these models, some
modification is made to GR to explain recent observations such as
the cosmic acceleration. These modifications include the adding of
extra dimensions like in the brane world models \cite{Brax03}, the
adding of a minimally or non-minimally coupled scalar field
\cite{STGrav}, or modifications of the underlying field equations by
either adding higher order corrections to the curvature \cite{DEfR}
or changing the equation of state \cite{NLEoS}. In general, these
modified theories of gravity have more complicated effective
evolution equations, and it can be more difficult to find exact
analytical solutions.

The implementation of the theory of dynamical systems (see for
example \cite{Dynamical,DSCosmo,Wainwright04}) has proven to be
useful to gain a qualitative understanding of a given class of
cosmological models. This dynamical systems approach does not
require the knowledge of any exact solutions. However, the
equilibrium points of the dynamical system correspond to the
interesting cosmological solutions. This approach helps identifying
exact solutions with special symmetries, which is particularly useful
when studying complicated field equations.

In recent times, this approach has been used to investigate
alternative theories of gravity such as Brans-Dicke theory
\cite{Burd88,Kolitch95,Kolitch96,Santos97,Holden98}, scalar-tensor
theories
\cite{Foster98,Amendola90,Billyard99,Holden00,Gunzig00,Gunzig01,Saa01,Faraoni05,Carloni07,Agarwal07},
and higher order gravity
\cite{Carloni05,Leach06,Abdel07,Clifton05,Barrow06c,Goheer07a,Amendola07a,Amendola07b,Cognola07,Carloni07a}.
It has also proven useful in theories with non-linear equations of
state \cite{Ananda06a,Ananda06b} and brane world models
\cite{Campos01a,Campos01b,Coley02a,Coley02b,Dunsby04,Goheer04}.

In this paper, we consider the various frameworks in which dynamical
systems theory can be applied to cosmology.  In section 2 we discuss
the characteristics of non-compact and compact state spaces in
general. We point out the advantages of compactifying the state
space, emphasising the aspect of static and bounce type solutions.
In section 3, we proceed to give the specific example of LRS Bianchi
I models in $R^n$--gravity and compare the results of
\cite{Goheer07a} and \cite{Leach06}, where compact and non--compact
expansion--normalised variables were used respectively. In section 4
we consider the flat Friedmann models in $R^n$--gravity and compare
the compact formalism of \cite{Goheer07a} to that of Clifton {\it et
al.} \cite{Clifton05}, where non-compact non--expansion--normalised
variables were used. We summarise our results in section 5, pointing
out the discrepancies between the three approaches \cite{Goheer07a},
\cite{Leach06} and \cite{Clifton05}.

%%%%%%%%%%%%%%%%%%%%%%%%%%%%%%%%%%%%%%%%%
\section{Choice of the state space}
%%%%%%%%%%%%%%%%%%%%%%%%%%%%%%%%%%%%%%%%%

In order to perform a dynamical systems analysis on homogeneous
cosmologies, one has to construct variables corresponding to the
kinematic quantities as well as a time variable that together define
an autonomous system of first-order differential equations. The
choice of variables depends on several physical considerations:
Firstly, one would like to study the cosmological behaviour close to
the initial singularity and the late time behaviour of the model.
Secondly, we want to study the effect of matter, shear and other
physical influences on the cosmological dynamics. Finally, we would
like to constrain the system by making use of observations such as
the cosmic microwave background.

The so--called Hubble-- or expansion--normalised variables together
with a Hubble--normalised time variable  \cite{Collins71} have been
used successfully to study important issues such the isotropisation
of cosmological models \cite{Collins73}. The state space defined by
the Hubble--normalised variables is compact for simple classes of
ever expanding models such as the open and flat FLRW models and the
spatially homogeneous Bianchi type I models in GR. In these cases, the
dynamical systems variables are bounded even close to the
cosmological singularity \cite{Wainwright04}. This is due to the
fact that these simple classes of cosmological models do not allow
for bouncing, recollapsing or static models, since there are no
contributions to the Friedman equation that would allow for the
Hubble--parameter to vanish.

As soon as there are additional degrees of freedom allowing $\Theta$
to pass through zero (e.g. the simple addition of positive spatial
curvature), the state space obtained from expansion--normalised
variables becomes non--compact. Note that even the time variable
becomes ill-defined in this case and needs to be used carefully (see
below). If the expansion normalised variables are unbounded, one has
to perform an additional analysis to study the equilibrium points at
infinity. This can be done using the well known Poincar\'{e}
projection \cite{Poincare,Perko}, where the points at infinity are
projected onto a unit sphere. These projected \equi points can then
be analysed in the standard way, i.e by considering small
perturbations around the points. However, it may still be difficult
to determine the stability of the \equi points at infinity.

Alternatively, one may break up the state space into compact
subsectors, where the dynamical systems and time variables are
normalised differently in each sector (see for e.g.
\cite{Goliath99}). The full state space is then obtained by pasting
the compact subsectors together. We will discuss these two methods
in the following subsections, highlighting the advantages and
disadvantages in this context.

%------------------------------------------------
\subsection{Non-compact state spaces and the Poincar\'{e} projection}
%------------------------------------------------

In cosmology it is not always straightforward to construct variables
defining a compact dynamical system associated with the class of
cosmological models of interest. This is especially true if one
considers more complicated theories such as modified theories of
gravity.  In many of the analyses of these types of theories, the
dynamical systems variables are not expansion--normalised and define
a non--compact state space
\cite{Burd88,Kolitch95,Kolitch96,Santos97,Holden98,Gunzig00,Gunzig01,Saa01,Faraoni05,Clifton05}.
These analyses make use of a conformal time, which places
restrictions on the ranges of physical quantities, such as the
energy density, Ricci scalar or scalar-field (see section 4). The
behaviour of the
 system at infinity can then studied using a Poincar\'{e} projection. In this framework,
 the \equi points at infinity represent the cosmological singularities such as
 initial singularities or other singularities where the scale factor,
scalar field or other variables of the system tend to zero.
Alternatively, the original physical variables may be used
\cite{Berkin90,Rendall06,Miritzis07}, but an asymptotic analysis is
still required to study initial singularities. Despite the
non-compactness of the state space constructed in these two
approaches, one may in principle study bouncing or recollapsing
behaviours as well as static solutions, since one does not normalise
with $\Theta$.

It is often useful to define expansion--normalised variables
together with a dimensionless, expansion--normalised time variable
in order to decouple the expansion rate from the remaining
propagation equations. This approach only yields a well--defined
time variable if we only study ever expanding or ever collapsing
models; a sign change in the expansion rate would make this time
variable non-monotonic. For the simple class of FLRW models in GR
for example, there are no bouncing or recollapsing models, and the
expanding or collapsing models can be studied separately in a
well--defined compact framework. In a more general scenario however,
static and recollapsing or bouncing solutions may occur, and one
would have to introduce a modified normalisation in order to define
a state space that includes these singularities.

In some cases it may be useful to employ expansion--normalised
variables, but it may not feasible to compactify the state space.
This is the case when e.g. only studying ever--expanding
cosmological models. Non-compact expansion--normalised variables
have been used successfully to study aspects of isotropisation in
higher order gravity models \cite{Leach06,Barrow06c}. As pointed out
above, the non--compact expansion--normalised state space can only
contain expanding (or, by time reversal, collapsing) solutions by
construction. In particular, one cannot easily study bounce
behaviours in this setup, since the expanding and collapsing
subspaces would have to be pasted together at infinity, which is
non-trivial. Furthermore, the time variable is ill-defined in this
limit and needs careful treatment.

%------------------------------------------------
\subsection{Compact state spaces}
%------------------------------------------------
As mentioned above, expansion--normalised variables define a compact
state space for certain simple classes of cosmologies such as the
class of flat Friedmann models in GR \cite{Collins71}. When e.g.
additionally allowing for positive spatial curvature however, this
behaviour breaks down even in GR. Formally, we have a negative
contribution to the Friedmann equation, allowing all the other
variables to become unbounded. Physically, the reason for the
non-compactness of the state space is that positive spatial
curvature allows for static and bouncing solutions which have
vanishing expansion rate at least at some point in time. At this
point in time, the simple Hubble--normalisation is ill-defined,
causing the expansion-normalised variables as well as the
expansion-normalised time to diverge.

In \cite{Goliath99} a simple formalism has been established to
compactify the state space: if any negative contribution to the
Friedmann equation is absorbed into the normalisation, one can
define compact expansion normalised variables. If there are any
quantities that may be positive or negative, one has to study each
option in a separate sector of the state space and obtain the full
state space by matching the various sectors along their common
boundaries. In particular, this choice of normalisation ensures that
the accordingly normalised time variable is well-defined and
monotonic, and the state space obtained in this way may include
static, bouncing and recollapsing models. This approach has been
successfully adapted to compactify the state space corresponding to
more complicated classes of cosmologies (see for example
\cite{Campos01a,Campos01b,Goheer04,Dunsby04}).

%%%%%%%%%%%%%%%%%%%%%%%%%%%%%%%%%%%%
\section{Example 1:  LRS Bianchi I cosmologies in $R^n$-gravity}
%%%%%%%%%%%%%%%%%%%%%%%%%%%%%%%%%%%%

 In this section we outline how the method discussed in
 \cite{Goheer07a} is used to construct a compact
 expansion--normalised state space for the simple class of LRS Bianchi I
 cosmologies. We then compare the results obtained in this framework
 to the results obtained using the non--compact expansion--normalised setup of \cite{Leach06}.
 We will express the \equi points and coordinates of the compact
analysis \cite{Goheer07a} with a tilde to distinguish them from the
corresponding points in the non-compact analysis \cite{Leach06}. We
end this section with a discussion of bouncing and recollapsing
models based on the compact framework.

The Bianchi I models in $R^n$--gravity are fully characterised by
the Raychaudhuri equation, the Gauss-Codazzi equation and the
Friedmann constraint:
\begin{eqnarray}
\dot{\Th} + \sfrac{1}{3}\,\Th^{2} + 2\sigma^2 -\frac{1}{2n}R -
(n-1)\frac{\dot{R}}{R}\Th+\frac{\mu}{nR^{n-1}} =0, \label{raychaudhuri}\\
 \dot{\sig}+\left(\Th+
(n-1)\frac{\dot{R}}{R}\right)\sigma=0, \label{sigma-dot}\\
\Th^{2}=3\sigma^2-3(n-1)\frac{\dot{R}}{R}\Th+\frac{3(n-1)}{2n}R+\frac{3\mu}{nR^{n-1}}.
\label{friedman}
\end{eqnarray}
Assuming that the standard matter behaves like a perfect fluid with
barotropic index $w$, the energy conservation equation is as usual
given by
\begin{equation}\label{cons:perfect}
\dot{\mu}+(1+w)\mu\Th=0\,.
\end{equation}
%-------------------------------------------
\subsection{Construction of the compact state space}
\begin{figure}[tbp]
\begin{center}
\epsfig{file=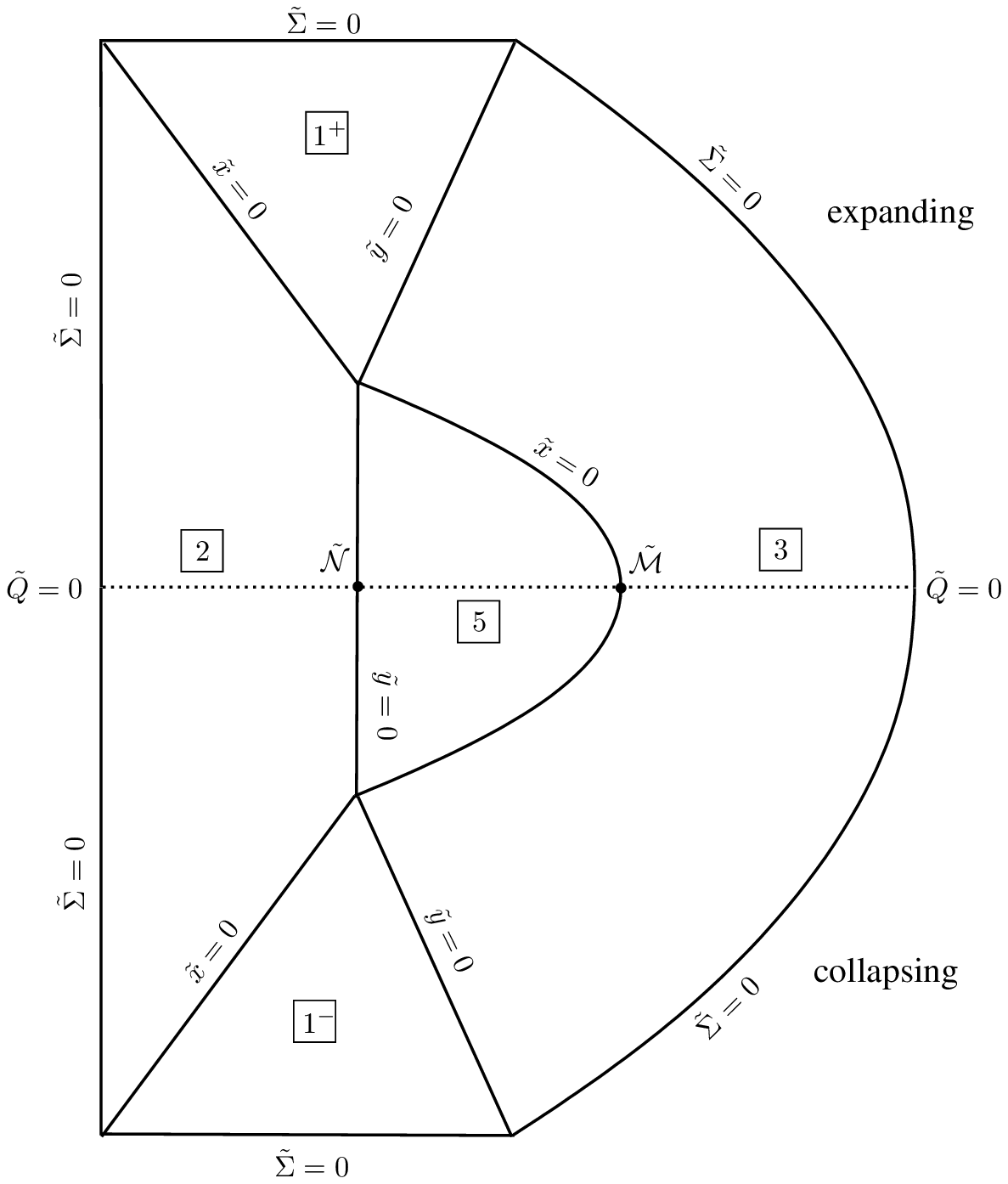, scale=0.6} \caption{Schematic construction
of the compact state space of the vacuum LRS Bianchi I models. The
sectors have been labeled (numbers in square boxes) according to
\cite{Goheer07a}. Note that the state space is symmetric around
$\Sigma=0$, so that this figure can represent $\Sigma\geq0$ or
$\Sigma\leq0$.}\label{fig:n_1}
\end{center}
\end{figure}

%-------------------------------------------
The LRS Bianchi I state space is compactified as discussed in detail
in \cite{Goheer07a}: we introduce the dynamical variables
\begin{eqnarray}
\label{def:var}
 \tilde{\Sigma} =\frac{\sqrt{3}\sigma}{D}\; , \quad &&\tilde{x} =
\frac{3\dot{R}\Th}{R D^2}(1-n)\; ,\quad \tilde{y} = \frac{3R}{2n D^2}(n-1)\;,\\
\tilde{z} = \frac{3\mu}{nR^{n-1}D^2}\; && \; ,\quad \quad \quad
\quad \tilde{Q}=\frac{\Th}{D} \,, \nonumber
\end{eqnarray}
and a dimensionless time variable $\tau$ defined by
\begin{equation}
\label{def:time-variable} \frac{d}{d\tau}\equiv
~'=\frac{1}{D}\frac{d}{dt}\,.
\end{equation}
The normalisation $D$ will be chosen such that it is strictly
positive at all times. As in GR, we have to explicitly exclude the
static flat isotropic vacuum cosmologies \cite{Goliath99}. We define
eight different sectors according to the possible signs of
$\tilde{x},~\tilde{y}$ and $\tilde{z}$\footnote{Note that the sign
of these quantities is independent of the exact choice of the
normalisation, since $D$ is real and enters quadratically.}. The
first sector is characterised by
$\tilde{x},~\tilde{y},~\tilde{z}\geq 0$.  In this sector, we can
simply choose $D=|\Theta|=\eps\Theta$, where $\eps=\pm 1$ is defined
to be the sign function of $\Theta$: $\eps=|\Theta|/\Theta$. The
Friedmann equation becomes
\begin{equation}\label{fried-S1}
1=\tilde{\Sigma}^2+\tilde{x}+\tilde{y}+\tilde{z}\,.
\end{equation}
By construction, all the contributions to the right hand side of
(\ref{fried-S1}) are positive, hence the variables
$\tilde{x},~\tilde{y}$ and $\tilde{z}$ have to take values in the
interval $[0,1]$, while $\tilde{\Sigma}$ must lie in $[-1,1]$. Note
that $\tilde{Q}=\eps=\pm 1$ is not a dynamical variable in this
sector only, where we have excluded $\Theta = 0$ as motivated above.
This means that we have to create two copies of this sector, one
corresponding to the expanding models ($\tilde{Q}=\eps=1$) and one
to corresponding to the collapsing models ($\tilde{Q}=\eps=-1$).
These two sectors are disconnected (see Figure \ref{fig:n_1}).

In the other sectors, we absorb any negative contributions to the
Friedmann equation into the normalisation. For example, if
$\tilde{x},~\tilde{y}>0$ and $\tilde{z}<0$ (as in sector 4 of
\cite{Goheer07a}), we define
$D=\sqrt{\Theta^2-\frac{3\mu}{nR^{n-1}}}$. Note that we may now
include the static and bouncing or recollapsing solutions with
$\Theta=0$ as long as there is matter present ($\mu\neq 0$). Again
we can express the Friedmann equation in terms of the
normalised variables (\ref{def:var}) and observe that $\tilde{z}$
does not explicitly appear, but all the other contributions enter
with a positive sign. This means that $\tilde{x},~\tilde{y}$ and
$\tilde{\Sigma}^2$ are positive and must take values in $[0,1]$. One
can easily show that $\tilde{z}$ is bounded by the interval $[-1,0]$
in this sector, and $\tilde{Q}$ lies in $[-1,1]$. The other sectors
are constructed by analogy (see \cite{Goheer07a} for details).

Note that in all the sectors other than the first one, $\tilde{Q}$
is a dynamical variable (taking values in $[-1,1]$) with the sign of
$\tilde{Q}$ corresponding to the sign of the Hubble factor. This
means that in these sectors, we naturally include both expanding and
collapsing models and do not have to artificially create two copies
of the sectors. Furthermore, we point out that in all sectors other
than the sectors 1 and 2, we can principally include static
solutions. The exclusion of static or bouncing/recollapsing models
in sector 1 has been explained above. Sector 2 is similar to sector
1 in the limit $\Theta=0$ because of the special way the variable
$x$ is defined: in this case the normalisation vanishes, and we
therefore have to exclude this case.

\begin{figure}[tbp]
\begin{center}
\epsfig{file=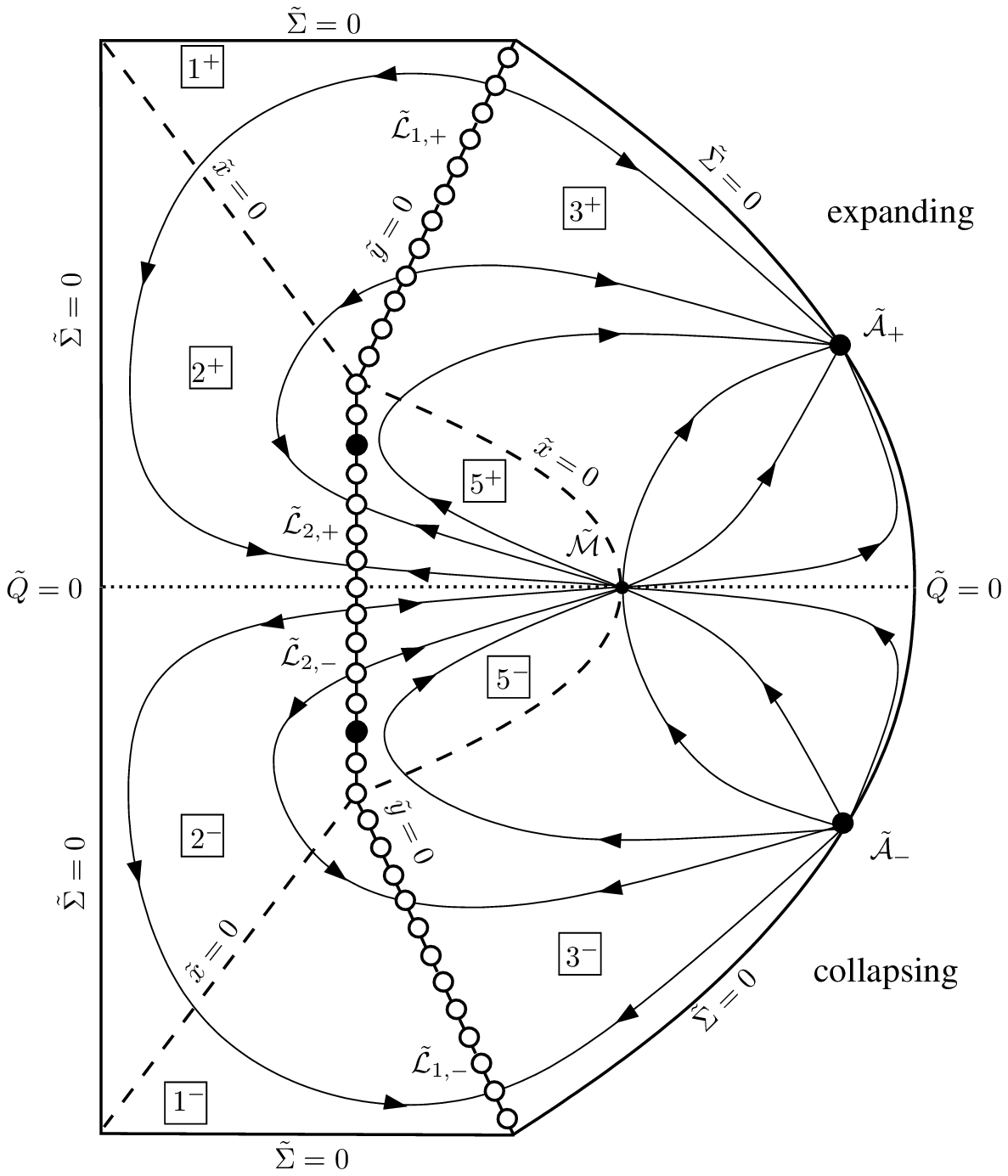, scale=0.6} \caption{Compactified state
space of the vacuum LRS Bianchi I models for $n\in (1/2,1)$. The
line of hollow circles corresponds to $\tilde{\mathcal{L}}_{1,2}$,
where the only points with corresponding cosmological solutions are
filled in solid. The dotted line represents the points with
$\tilde{Q}=0$, but only the point $\tilde{\mathcal{M}}$ with
$\tilde{x}=\tilde{Q}=0$ and the point on the line
$\tilde{\mathcal{L}}_{2}$ with $\tilde{y}=\tilde{Q}=0$ (referred to
as $\tilde{\mathcal{N}}$ in the text) may have corresponding
solutions to the underlying field equations. We emphasise that
$\tilde{\mathcal{M}}$ is not an \equi point - it is highlighted
because it represents the only point where orbits in this subspace
can cross between the expanding and collapsing sectors. See text for
more details.}\label{fig:n_a}
\end{center}
\end{figure}

\begin{figure}[tbp]
\begin{center}
\epsfig{file=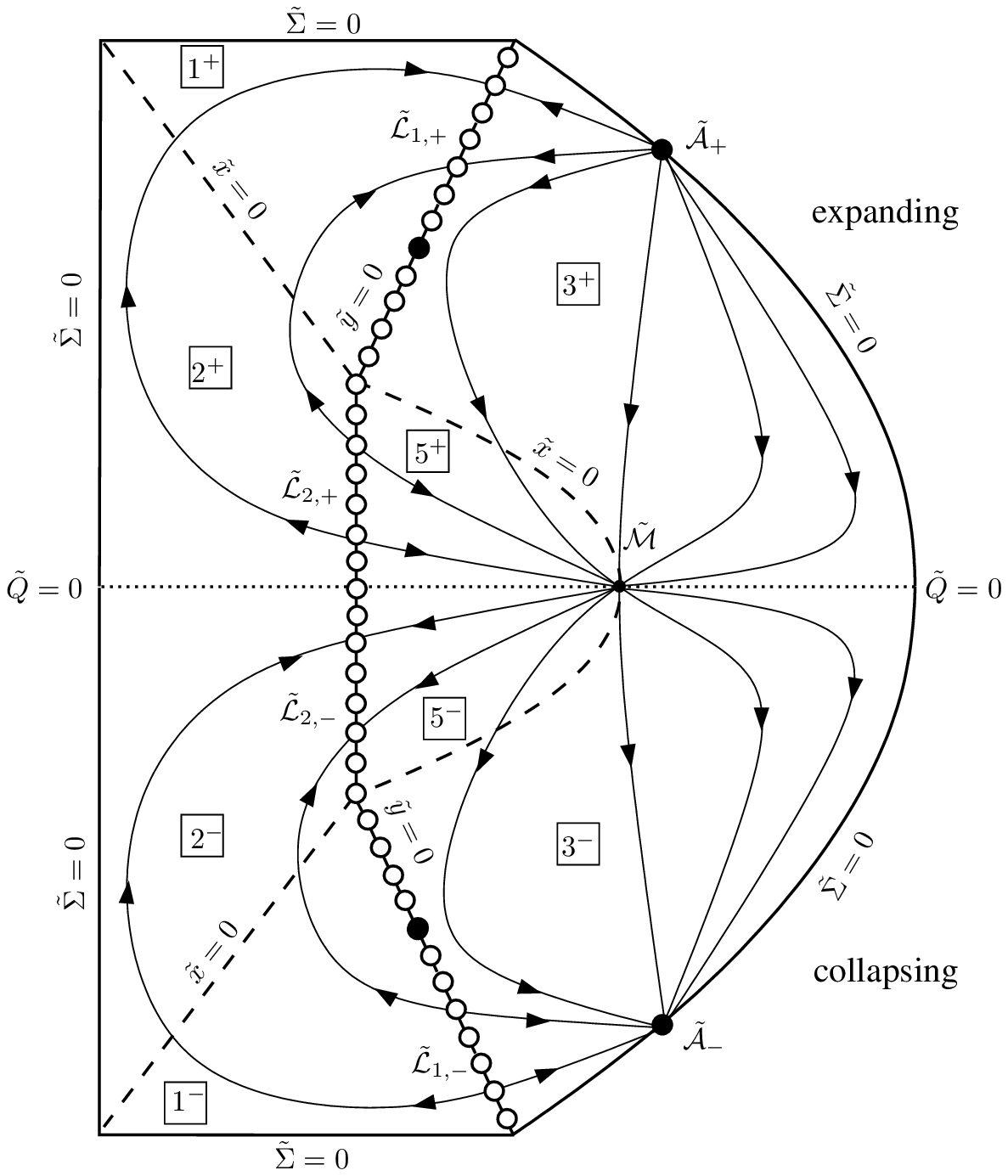, scale=0.6} \caption{Compactified state
space of the vacuum LRS Bianchi I models for $n\in (1,5/4)$. See
caption of Figure \ref{fig:n_a} for more details.}\label{fig:n_b}
\end{center}
\end{figure}

The full state space is obtained by matching the various sectors
along their common boundaries defined by
$\tilde{x},~\tilde{y},~\tilde{z}=0$. For simplicity, we will first
address the vacuum subspace ($\tilde{z}=0$). This space consists of
four 2-dimensional compact sectors corresponding to the sign of the
variables $\tilde{x}$ and $\tilde{y}$. As discussed above, we have
to create the two copies of the first sector (labeled $1^+$ and
$1^-$) corresponding to the disconnected expanding and collapsing
parts respectively. The full state space is then composed of five
different pieces as depicted schematically in Figure \ref{fig:n_1}.
Strictly speaking we have to exclude the points with $\tilde{Q}=0$
and $\tilde{x},~\tilde{y}\neq 0$, since $\tilde{Q}=0$ implies
$\Theta=0$ which in turn implies $\tilde{x}=0$ unless $R=0$. This is
indicated with a dotted line in Figures \ref{fig:n_1},
\ref{fig:n_a}, \ref{fig:n_b} and \ref{fig:n_c}, showing that the
$\tilde{Q}=0$ plane may only be crossed at the points with
$\tilde{x}=0$ or $\tilde{y}=0$. We will label these points
$\tilde{\mathcal{M}}$ and $\tilde{\mathcal{N}}$ respectively.

The points with $\tilde{y}=0$ have to be treated with caution: these
points necessarily have vanishing Ricci scalar $R$ and the
corresponding cosmological solutions can only be discussed in the
limit $R\rightarrow 0$. This issue is addressed in detail in
\cite{Goheer07a}, where it was found that there only exist solutions
corresponding to these points for very special values of $n$. The
same issue applies to point $\tilde{\mathcal{N}}$, which is a
degenerate point as discussed is section 4 below.

The state space corresponding to the matter case is 3--dimensional
and consists of eight separate sectors. It is straightforward to
construct by analogy with the vacuum case, but harder to present in
a graphic visualisation because of the higher dimensionality of the
state space. We therefore omit a graphic representation of the
matter state space.

We point out that unlike in the vacuum case, where $\Theta=0$ was
only allowed at the single points $\tilde{\mathcal{M}}$ and
$\tilde{\mathcal{N}}$, the matter case allows for one additional
degree of freedom. In this case static or bouncing/recollapsing
models must pass through the 1--dimensional lines extending
$\tilde{\mathcal{M}}$ and $\tilde{\mathcal{N}}$ along the
$z$--direction, and can therefore occur at for a wider range of
variables. This is of course due to the fact that the curvature term
coupled to the matter contribution can counterbalance the other
terms in the Friedmann equation.

%-------------------------------------------
\subsection{Comparison of equilibrium points}
%-------------------------------------------
We first look at the vacuum \equi points found in \cite{Leach06} and
in  the LRS  Bianchi I state subspace of \cite{Goheer07a}. Since the
former paper only considered expanding models, we restrict ourselves
to the expanding subset of the compact LRS Bianchi I state subspace
of \cite{Goheer07a} in this comparison.

In the non-compact analysis \cite{Leach06}, one Friedmann-like \equi
point $\mathcal{A}$, a line of \equi points $\mathcal{L}_1$
corresponding to Bianchi I models, and four asymptotic \equi points
$\mathcal{A}_\infty$, $\mathcal{B}_\infty$, $\mathcal{C}_\infty$ and
$\mathcal{D}_\infty$ were found. The coordinates of $\mathcal{A}$
diverge as $n\rightarrow 1/2$. This means that for this bifurcation
value the point moves to infinity, where it merges with the
asymptotic \equi point $\mathcal{B}_\infty$ if $n \rightarrow
{1}/{2}^-$ and with $\mathcal{C}_\infty$ if $n \rightarrow
{1}/{2}^+$. The asymptotic point $\mathcal{A}_\infty$ is the
``endpoint" of $\mathcal{L}_{1}$ at infinity. The four \equi points
at infinity occur for all values of $n$. Note that for $n=1/2$ a
bifurcation occurs, where the isolated asymptotic \equi points turn
into a ring of \equi points at infinity. This means that for this
value of $n$, asymptotic \equi points occur at all angles. This
bifurcation was not considered in \cite{Leach06}.

In the compact analysis \cite{Goheer07a}, one Friedmann-like \equi
point $\tilde{\mathcal{A}}$ and two Bianchi I lines of \equi points
$\tilde{\mathcal{L}}_{1}$ and $\tilde{\mathcal{L}}_{2}$ were found
in the flat vacuum subspace explored here. Note that the two lines
are in fact the same but for different signs of the variable
$\tilde{x}$ (see below).

Table~\ref{points:vac} summarises the \equi points from the compact
analysis \cite{Goheer07a}  and the corresponding counterparts in the
non-compact analysis \cite{Leach06}. We can see that the finite
\equi points in \cite{Leach06} correspond to the similarly labeled
ones in \cite{Goheer07a} for all values $n$, even for the
bifurcation values of $n$ for which the finite points in
\cite{Leach06} move to infinity. We note that the asymptotic points
$\mathcal{B}_\infty$ and $\mathcal{C}_\infty$ only have analogs in
\cite{Goheer07a} for the bifurcation value $n=1/2$. The line
$\mathcal{L}_{1}$ in \cite{Leach06} corresponds to
$\tilde{\mathcal{L}}_{1}$ in \cite{Goheer07a} for $\Sigma_* \in
[0,1]$ and to $\tilde{\mathcal{L}}_{2}$ for $\Sigma_{*}>1$, where
$\Sigma_{*}$ parametrises the line $\mathcal{L}_{1}$.
$\mathcal{A}_\infty$ corresponds to the single static
($\tilde{Q}_*=0$) point on $\tilde{\mathcal{L}}_{2}$ in
\cite{Goheer07a}, labeled $\tilde{\mathcal{N}}$ in this section.

 The \equi
point $\mathcal{D}_\infty$ in \cite{Leach06} corresponds to the
point $\tilde{\mathcal{M}}$ in the compact analysis as noted in
Figures \ref{fig:n_a} and \ref{fig:n_b}. Note that
$\tilde{\mathcal{M}}$ is not an \equi point in \cite{Goheer07a}; it
only appears to be an \equi point in the non-compact analysis
because in this case only the expanding half of the full state space
was studied. When including the collapsing part of the state space
as done in \cite{Goheer07a}, it becomes clear that
$\tilde{\mathcal{M}}$ merely denotes the point at which orbits may
cross between the expanding and contracting parts of the state
space.

\begin{table}[tbp] \centering
\caption{Correspondence between the equilibrium points of the vacuum
LRS Bianchi I state space in the compact \cite{Goheer07a} and
non--compact \cite{Leach06} analysis. The last column states for
which parameter values the correspondence occurs. Note that
$\tilde{\mathcal{M}}$ is not an \equi point in the compact analysis;
we have only included the last row for completeness.}
\begin{tabular}{lll}
& & \\ \br  Compact & Non-compact & parameter constraint \\
\br $\tilde{\mathcal{A}}_+$ & $\left\{\begin{tabular}{l}
$\mathcal{A}$
\\ $\mathcal{B}_\infty$ \\ $\mathcal{C}_\infty$ \end{tabular}\right.$ &
\begin{tabular}{l} all $n$ \\ $n \rightarrow \sfrac{1}{2}^-$ \\
$n \rightarrow \sfrac{1}{2}^+$\end{tabular} \\
& & \\$\tilde{\mathcal{L}}_{1,+}$ &  \ \ \ \ $\mathcal{L}_1$ & \ \
all $n$,~$\Sigma_* \in [0,1]$
\\
& & \\
$\tilde{\mathcal{L}}_{2,+}$ & $\left\{\begin{tabular}{l}
$\mathcal{L}_1$
 \\ $\mathcal{A}_\infty$
\end{tabular}\right.$ & \begin{tabular}{l} all $n$,~$\Sigma_*\in (1,\infty)$ \\
all $n$,~$\Sigma_*\rightarrow \infty$ ($\tilde{Q}_*=0$) \end{tabular} \\ & & \\
($\tilde{\mathcal{M}}$)  &  \ \ \ \ $\mathcal{D}_\infty$ &  \ \ all $n$\\
\br
\end{tabular}\label{points:vac}
\end{table}

We now consider the matter equilibrium points. In the non-compact
analysis \cite{Leach06} three finite isotropic points were found:
the vacuum point $\mathcal{A}$ and two non-vacuum points
$\mathcal{B}$ and $\mathcal{C}$. Furthermore, the vacuum Bianchi I
line of points $\mathcal{L}_1$ was recovered. There were five
asymptotic \equi points $\mathcal{A}_\infty$, $\mathcal{B}_\infty$,
$\mathcal{C}_\infty$, $\mathcal{D}_\infty$ and $\mathcal{E}_\infty$,
and a line of \equi points denoted $\mathcal{L}_\infty$. The
coordinates of $\mathcal{A}$ diverge when $n \rightarrow {1}/{2}$;
in this case $\mathcal{A}$ merges with $\mathcal{D}_\infty$ when $n
\rightarrow {1}/{2}^-$ and with $\mathcal{E}_\infty$ when $n
\rightarrow {1}/{2}^+$. Similarly, the coordinates of \equi point
$\mathcal{C}$ approach infinity when $n \rightarrow0$: $\mathcal{C}$
merges with $\mathcal{A}_\infty$ when $n \rightarrow 0^-$ and
 with $\mathcal{B}_\infty$ when $n \rightarrow0^+$.
Point $\mathcal{B}$ on the other hand remains a finite \equi point
for all values of $n$. As pointed out in the vacuum case,
$\mathcal{C}_\infty$ is the ``endpoint" of $\mathcal{L}_{1}$ at
infinity with $\Sigma_\ast\rightarrow \infty$. As in the vacuum
case, there is ring of asymptotic fixed points in the $z=0$ plane at
the bifurcation value $n=1/2$. Furthermore, there is a ring of \equi
points in the $\Sigma=0$ plane at the bifurcation value $n=0$. These
bifurcations have not been noted in \cite{Leach06}.

In the compact analysis \cite{Goheer07a}, the three isotropic points
$\tilde{\mathcal{A}}$, $\tilde{\mathcal{B}}$ and
$\tilde{\mathcal{C}}$ and the two vacuum  Bianchi I lines of \equi
points $\tilde{\mathcal{L}}_{1}$ and $\tilde{\mathcal{L}}_{2}$ were
found. As in the vacuum case, we can see the correspondence between
the \equi points in \cite{Goheer07a} and the ones in \cite{Leach06},
where the finite \equi points may move to infinity for certain
values of $n$. We have summarised these results in
Table~\ref{points:mat}.

The line  $\mathcal{L}_\infty$ is the higher dimensional matter
analog of point $\mathcal{D}_\infty$ from the  vacuum analysis: the
counterpart of $\mathcal{L}_\infty$ in \cite{Goheer07a}  is not a
line of \equi points. $\mathcal{L}_{\infty}$ only appears as a line
of \equi points in \cite{Leach06} because the collapsing part of the
state space is not included (see above).

\begin{table}[tbp] \centering
\caption{Correspondence between the equilibrium points of the LRS
Bianchi I state space with matter in the compact \cite{Goheer07a}
and non--compact \cite{Leach06} analysis. The last column states for
which parameter values the correspondence occurs. Here the extension
of $\tilde{\mathcal{M}}$ is the 1--dimensional generalisation of
point $\tilde{\mathcal{M}}$ for the matter case (see text for
details), which is not an \equi point in the compact analysis; we
have only included the last row for completeness.}
\begin{tabular}{lll}
& & \\ \br  Compact & Non-compact & parameter constraints \\
\br $\tilde{\mathcal{A}}_+$ & $\left\{\begin{tabular}{l}
$\mathcal{A}$
\\ $\mathcal{D}_\infty$ \\ $\mathcal{E}_\infty$ \end{tabular}\right.$ &
\begin{tabular}{l} all $n$\\ $n \rightarrow \sfrac{1}{2}^-$ \\
$n \rightarrow \sfrac{1}{2}^+$\end{tabular} \\
& & \\
$\tilde{\mathcal{B}}_+$ &  \ \ \ \ $\mathcal{B}$ &  \ \ all $n$ \\
& & \\
$\tilde{\mathcal{C}}_+$ & $\left\{\begin{tabular}{l} $\mathcal{C}$
\\ $\mathcal{A}_\infty$ \\ $\mathcal{B}_\infty$ \end{tabular}\right.$ &
\begin{tabular}{l} all $n$\\ $n \rightarrow 0^-$ \\
$n \rightarrow 0^+$\end{tabular} \\
& & \\
$\tilde{\mathcal{L}}_{1,+}$ &  \ \ \ \ $\mathcal{L}_1$ & \ \ all
$n$,~$\Sigma_* \in [0,1]$
\\
& & \\
$\tilde{\mathcal{L}}_{2,+}$ & $\left\{\begin{tabular}{l}
$\mathcal{L}_1$
 \\ $\mathcal{C}_\infty$
\end{tabular}\right.$ & \begin{tabular}{l} all $n$,~$\Sigma_*\in (1,\infty)$ \\
all $n$,~$\Sigma_*\rightarrow \infty$ ($\tilde{Q}_*=0$) \end{tabular} \\ & & \\
(extension of $\tilde{\mathcal{M}}$) &  \ \ \ \  $\mathcal{L}_\infty$ &  \ \ all $n$\\
\br
\end{tabular}\label{points:mat}
\end{table}

%-------------------------------------------
\subsection{Solutions and stability}
%-------------------------------------------

In both \cite{Leach06} and \cite{Goheer07a}, the exact solutions to
the field equations (\ref{raychaudhuri})-(\ref{friedman})
corresponding to each \equi point were derived.

The solutions to the finite points in \cite{Leach06} are the same as
the ones obtained in \cite{Goheer07a} for their counterparts in the
compact analysis except for the points with $y=0$, which are very
special, since they necessarily have vanishing Ricci scalar $R$ and
their solutions can only be obtained in a careful limiting
procedure. This is discussed in great detail in \cite{Goheer07a},
where it was found that these points only have corresponding
solutions for very special values of $n$. In the LRS Bianchi I state
space discussed here, the point $\tilde{\mathcal{B}}$ and the lines
$\tilde{\mathcal{L}}_{1,2}$ have vanishing Ricci scalar. As
discussed in \cite{Goheer07a}, $\tilde{\mathcal{B}}$ only has a
solution for the bifurcation value $n={5}/{4}$ and $w={2}/{3}$, and
 $\tilde{\mathcal{C}}_\pm$ only admits a solution for
$n\in(1,N_+)$, where we abbreviate
$N_\pm=\frac{1}{4(4+3w)}\left(13+9w \pm \sqrt{9 w ^2+66 w
+73}\right)$. Only two points on $\tilde{\mathcal{L}}_{1,2}$ have
corresponding cosmological solutions. These points are marked with
solid circles in Figures \ref{fig:n_a} and \ref{fig:n_b}. This issue
was not addressed in \cite{Leach06}, where the authors did not solve
the full set of field equations to obtain the exact solutions.
However, this is not a problem caused by the use of non-compact
variables; the results of \cite{Goheer07a} can be recovered using
the setup of \cite{Leach06} if one carefully solves for \emph{all}
cosmological variables (including $R$).

The solutions for the asymptotic \equi points in \cite{Leach06}
differ from the ones obtained in \cite{Goheer07a} for the
corresponding \equi points. In \cite{Leach06}, the solutions
corresponding to the asymptotic vacuum points $\mathcal{B}_\infty$,
$\mathcal{C}_\infty$, $\mathcal{D}_\infty$ and the asymptotic matter
points $\mathcal{A}_\infty$, $\mathcal{B}_\infty$,
$\mathcal{D}_\infty$, $\mathcal{E}_\infty$ were all de Sitter like.
In \cite{Goheer07a} on the other hand, it was shown that there are
no solutions to the corresponding \equi points.

A careful analysis shows that the stationary solutions in
\cite{Leach06} are not valid, since they cannot simultaneously
satisfy the evolution equations and the definitions of the
dimensionless expansion normalised variables in this limit. Note
that these solutions can satisfy the coordinates of the asymptotic
\equi points in the special static case. However, the static models
do not satisfy all the original field equations for this class of
models and therefore are not solutions as shown in \cite{Goheer07a}.
This was not investigated in \cite{Leach06}.

The solutions for the vacuum point $\mathcal{A}_\infty$  and the
matter point $\mathcal{C}_\infty$ given in \cite{Leach06} were
static in the appropriate limit $\Sigma_*\rightarrow \infty$. In
\cite{Goheer07a} however it is shown that the static models do not
satisfy all the evolution equations and therefore do not present
cosmological solutions. In this sense, we call all these \equi
points `unphysical'.

We conclude that while \cite{Leach06} and \cite{Goheer07a} found the
same solutions for the finite \equi points with $y\neq 0$ in
\cite{Leach06}, there was disagreement with the solutions
corresponding to the asymptotic points in \cite{Leach06} and the
points with $y=0$. This discrepancy arises from the fact that the
non--compact framework is much more complicated, so that it was not
noticed that the given solutions indeed do not simultaneously
satisfy the original equations and the coordinates of the
(asymptotic) \equi points.

The nature of the \equi points remains unchanged in both formalisms,
even though the time variable in \cite{Leach06} is strictly speaking
not well-defined at infinity. The reason for this agreement is that
we study perturbations away from the \equi points, i.e. strictly
speaking we never reach infinity when studying the eigenvalues. As
long as the given point is actually an \emph{ \equi} point, the
results from the expanding sector alone reflect the dynamical nature
of the point in the entire state space correctly. However, we
emphasise again that $\mathcal{D}_\infty$ and $\mathcal{L}_\infty$
in \cite{Leach06} are in fact not \equi points in \cite{Goheer07a}.
As explained in the previous subsection, they only appear to be
fixed points in the non--compact dynamical system. The stability of
$\mathcal{D}_\infty$ and $\mathcal{L}_\infty$ only indicates the
direction of the orbits in the expanding part of the compact state
space in \cite{Goheer07a}.

%-------------------------------------------
\subsection{Bounce behaviours}
%-------------------------------------------

As motivated above, the non-compact expansion normalised variables
are not suitable to study trajectories that correspond to
recollapsing or bouncing cosmologies \footnote{Static models may be
studied if one carefully takes into consideration that they have to
be analysed separately in the two copies corresponding to expanding
and collapsing models.}. We therefore only discuss this issue in the
context of \cite{Goheer07a}.

We start with the vacuum case: from the Raychaudhuri and Friedmann
equations, one can see that there can be bouncing solutions for
$n\in(1/2,1)$ and recollapsing solutions for $n\in(0,1/2)$ or $n>1$.
This is reflected in Figures \ref{fig:n_a} and \ref{fig:n_b}: we can
see that there are trajectories corresponding to bouncing solutions
for $n\in (1/2,1)$ and to recollapsing solutions for $n\in (1,5/4)$.
In both cases, the bouncing or recollapsing trajectories have to go
through the point $\tilde{\mathcal{M}}$, which is characterised by
$\tilde{x}=\tilde{Q}=0$. The existence of these bouncing or
recollapsing solutions has been confirmed by a numerical analysis.
For $n>1$, the recollapsing models have a negative Ricci tensor $R$,
while the bouncing or recollapsing models for $n<1$ have a positive
value of $R$.

In the matter case, there is one more degree of freedom. Any
bouncing or recollapsing solution must now pass through the
1-dimensional extension of point $\tilde{\mathcal{M}}$ in the
$z$--direction. This means it is easier to achieve bouncing or
recollapsing behaviour in the matter case. In particular, there can
be a bounce or recollapse even if $\tilde{y}>0$ (if
$\tilde{z}<-\tilde{y}$). Note that even though at first sight we
also expect bouncing behaviours through the 1--dimensional extension
of  $\tilde{\mathcal{N}}$, this line in fact corresponds to
degenerate cosmological models, and orbits approaching the line can
never reach or cross it as explained in detail in the next example.

%%%%%%%%%%%%%%%%%%%%%%%%%%%%%%%%%%%%
\section{Example 2:  Flat Friedmann cosmologies in $R^n$-gravity}
%%%%%%%%%%%%%%%%%%%%%%%%%%%%%%%%%%%%

In this section we consider the flat FLRW models with matter. We
will compare the results of \cite{Goheer07a} with the results of
Clifton {\it et al.} \cite{Clifton05}, where non-compact
non--expansion--normalised variables were used.

We briefly summarise the approach used by \cite{Clifton05} (which
follows \cite{Burd88,Holden98}): A conformal time coordinate
\begin{equation}\label{conftime}
d\tau=\sqrt{\frac{8\pi \mu}{3 R^{n-1}}}\;dt,
\end{equation}
and the dynamical variables
\begin{equation}\label{NNC:var}
X=\frac{R'}{R} \ \ \ \ {\rm and} \ \ \ \ Y=\frac{a'}{a},
\end{equation}
are introduced, where the primes here denote differentiation with
respect to $\tau$. Note that $\tau$ is only valid when $R^{n-1}$ is
a positive real root of $R$, which restricts the ranges of $R$ and
$n$.

Using the evolution equations \rf{raychaudhuri} and \rf{friedman}
(with $\sigma=0$),  an autonomous set of first order differential
equations for the variables $X$ and $Y$ is derived. This system is
non--compact, and is then analysed using standard dynamical systems
methods together with the Poincar\'{e} projection. Note that this
approach does not exclude models with $\Theta=0$, which allowed the
authors of \cite{Clifton05} to study static and bouncing or
recollapsing models.

The class of flat FLRW cosmologies is the isotropic subspace of the
class of LRS Bianchi I models studied in the previous section. We
can therefore simply take over the framework from \cite{Goheer07a}
as outlined in section 3. The \equi points for the FLRW state space
are simply the isotropic \equi points from the previous example.
Note that, unlike in the previous example, we now include both the
expanding and collapsing sectors in order to compare to
\cite{Clifton05}.

%-------------------------------------------
\subsection{Comparison of equilibrium points}
%-------------------------------------------

The analysis in \cite{Clifton05} yielded the two pairs of finite
\equi points $1,2$ and $3,4$. Furthermore, three pairs of \equi
points at infinity were found: a pair of static points $5,6$ and the
two pairs $7,8$ and $9,10$ with power law solutions. The odd and
even numbers in each pair correspond to expanding and collapsing
models depending on the value of $n$. Note that points $1,2$ only
have real coordinates for $n>0$ and $w<2/3$, while $3,4$ only have
real coordinates for $n\in(N_-,N_+)$, where
$N_\pm=\frac{1}{4(4+3w)}\left(13+9w \pm \sqrt{9 w ^2+66 w
+73}\right)$.  The pair $1,2$ merges with $7,8$ for $n=0$ or
$w=\sfrac{2}{3}$. Pair $3,4$ merges with $5,6$ for $n=0$, and $9,10$
merges with $5,6$ for $n=\sfrac{1}{2}$.

The compact analysis \cite{Goheer07a} yields three flat Friedmann
points (see section 3): $\tilde{\mathcal{A}}_\pm$,
$\tilde{\mathcal{B}}_\pm$ and $\tilde{\mathcal{C}}_\pm$, where the
expanding solutions are indicated by a plus and collapsing ones by a
minus subscript. As noted in the previous section,
$\tilde{\mathcal{B}}_\pm$ only admits a solution at the bifurcation
$n=5/4$ and $w=2/3$, while $\tilde{\mathcal{C}}_\pm$ only has a
cosmological solution for $n\in(1,N_+)$ and $w>-1$.

In Table \ref{points:mat2} we summarise the \equi points from
\cite{Goheer07a} and the corresponding counterparts in the analysis
of \cite{Clifton05}. We note that the two matter solutions
$\tilde{\mathcal{B}}_\pm$ and $\tilde{\mathcal{C}}_\pm$ in
\cite{Goheer07a} correspond to the finite \equi points $1,2$ and
$3,4$ in \cite{Clifton05}, while the vacuum \equi point
$\tilde{\mathcal{A}}_\pm$ in \cite{Goheer07a} corresponds to the
\equi points at infinity in \cite{Clifton05}. This is due to the
choice of coordinates \rf{NNC:var}, which diverge for
$\mu\rightarrow 0$.

We now discuss in detail for which parameter values the
correspondence between the \equi points occurs. We find that the
expanding (collapsing) point $\tilde{\mathcal{A}}_+$
($\tilde{\mathcal{A}}_-$) corresponds to $9$ ($10$) for
$n\in(1/2,1)$ or $n>2$,  and to point $10$ ($9$) for $n\in(0,1/2)$
or $n\in(1,2)$. There is no dependence on the equation of state
parameter $w$ in this case, since the point
$\tilde{\mathcal{A}}_\pm$ corresponds to a vacuum solution. The
matter point $\tilde{\mathcal{B}}_+$ ($\tilde{\mathcal{B}}_-$)
corresponds to point $1$ ($2$) for all $n>0$ provided
$w<\sfrac{2}{3}$, while for $w=2/3$ point $\tilde{\mathcal{B}}_+$
($\tilde{\mathcal{B}}_-$) corresponds to point $7$ ($8$) when $n>1$
and to point $8$ ($7$) when $n\in(0,1)$. Note that the matter point
$\tilde{\mathcal{C}}_+$ ($\tilde{\mathcal{C}}_-$) corresponds to $3$
($4$) over the entire allowed range of $n$.

As in the previous example, an \equi point at infinity in the
non--compact analysis (here $5,6$ or $7,8$) only has an analog in
the compact framework for the specific bifurcation values (of $n$
and in this case $w$) for which a finite \equi point moves to
infinity and merges with the respective asymptotic point.

We now give special consideration to the points $5,6$. We observe
that the two points $5$ and $6$ have the single analog
$\tilde{\mathcal{N}}$ in the compact analysis, which is not an \equi
point in \cite{Goheer07a}. The reason for this discrepancy is the
following: the points $5,6$ correspond to the limit $R\rightarrow 0$
(see equation (14) in \cite{Clifton05}). As pointed out in
\cite{Carloni05}, the plane $R=0$ is invariant, so that orbits
approaching this plane must turn around. Assuming $R$ starts out
positive and approaches zero, it is clear that the limit from the
left corresponds to $R'<0$ and $X\rightarrow -\infty$, while the
limit from the right corresponds $R'>0$ and $X\rightarrow \infty$.
Thus $5,6$  are not \equi points in the compact analysis: while they
appear as sink and source respectively in \cite{Clifton05}, they
merge into the single transitory point $\tilde{\mathcal{N}}$ in
\cite{Goheer07a}, similar to the case of $\mathcal{D}_\infty$ above.

However, this point $\tilde{\mathcal{N}}$ represents a singular
state: here $R=\Th=\dot{\Th}=0$, that means the field equations
break down and can only be studied in a careful limiting procedure
(see \cite{Goheer07a}). In particular, orbits approaching
$\tilde{\mathcal{N}}$ asymptotically slow down and never  reach or
pass through the point. In this sense we recover the results of
\cite{Clifton05}: even though the two disconnected points $5,6$ have
merged into the single point $\tilde{\mathcal{N}}$, no orbits can
pass through $\tilde{\mathcal{N}}$ and therefore the qualitative
result from \cite{Clifton05} is maintained.

\begin{table}[tbp] \centering
\caption{Correspondence between the equilibrium points in the state
space of the flat FLRW models with matter in the compact
\cite{Goheer07a} and non--compact \cite{Clifton05} analysis. The
plus-minus subscript indicates expanding and collapsing solutions
respectively, and the last column states for which parameter values
the correspondence occurs. We have abbreviated
$N_\pm=\frac{1}{4(4+3w)}\left(13+9w \pm \sqrt{9 w ^2+66 w
+73}\right)$.}
\begin{tabular}{lll}
& & \\ \br  Compact & Non-compact & parameter constraints \\
\br $\tilde{\mathcal{A}}_\pm$ & $\left\{\begin{tabular}{l} $5,6$
\\ $9,10$  \end{tabular}\right.$ &
\begin{tabular}{l} $n\rightarrow\sfrac{1}{2}$ \\ all $n$ \end{tabular} \\
& & \\
$\tilde{\mathcal{B}}_\pm$ &$\left\{\begin{tabular}{l} $1,2$
\\ $7,8$  \end{tabular}\right.$ &
\begin{tabular}{l} $w< \sfrac{2}{3}$ \\ $w=\sfrac{2}{3}$ \end{tabular} \\
& & \\
$\tilde{\mathcal{C}}_\pm$ & $\left\{\begin{tabular}{l} $3,4$
\\ $5,6$  \end{tabular}\right.$ &
\begin{tabular}{l} $n\in(N_-,N_+)$ \\
$n \rightarrow 0$\end{tabular} \\
& & \\
($\tilde{\mathcal{N}}$) & \ \ \ \ $5,6$ & \ \ all $n$ \\
\br
\end{tabular}\label{points:mat2}
\end{table}

%-------------------------------------------
\subsection{Solutions and stability}
%-------------------------------------------

The solutions given in \cite{Clifton05} have corresponding solutions
in \cite{Goheer07a} but only for specific values of the parameters
(see section 3.3). The solutions for the points $7,8$ are the same
as those found for $\tilde{\mathcal{B}}_\pm$ when $n=5/4$ and
$w=2/3$. In \cite{Clifton05}, points $1,2$ have the same solutions
as $7,8$ but they have no corresponding solutions in
\cite{Goheer07a}. Points $3,4$ have the same solution as
$\tilde{\mathcal{C}}_\pm$, and points $9,10$ have the same solutions
as $\tilde{\mathcal{A}}_\pm$. The static solutions for points $5,6$
given in \cite{Clifton05} do not satisfy all the evolution equations
and therefore are strictly speaking no exact solutions. This result
was also found in \cite{Goheer07a} for the points
$\tilde{\mathcal{A}}_\pm$ for $n\rightarrow 1/2$ and
$\tilde{\mathcal{C}}_\pm$ for $n\rightarrow 0$, which again reflects
the correspondence between the points in the respective limits.

The nature of the \equi points in \cite{Goheer07a} agrees with the
stability properties of the corresponding points in
\cite{Clifton05}. We note that the \equi points at infinity, $5,6$
and $7,8$, only have corresponding points in \cite{Goheer07a} for
specific values of $n$ and $w$ respectively. These parameter values
correspond to the bifurcations where the stability of the \equi
points changes and were not analysed in detail in \cite{Goheer07a}.
We therefore do not compare the two formalisms in this case.

%-------------------------------------------
\subsection{Bounce behaviours}
%-------------------------------------------
Unlike in the expansion--normalised non--compact analysis
\cite{Leach06} studied in the previous example, bouncing or
recollapsing solutions can be investigated the non-compact formalism
of \cite{Clifton05}, where specific examples were given: it was
shown that for matter with $w=0$ recollapsing solutions occur for
$n=1.1$ and bouncing solutions occur for $n=0.9$. \footnote{Exact
solutions corresponding to these bounces are given in
\cite{Clifton07}.}

We confirm this in our compact analysis, where as in the LRS Bianchi
I case, there are bouncing orbits through point
$\tilde{\mathcal{M}}$ for $n\in(1/2,1)$ and recollapsing orbits
through $\tilde{\mathcal{M}}$ for $n>1$. For $n\in(1/2,1)$, only
point $\tilde{\mathcal{A}}_\pm$ is physical, and so the physically
relevant behaviour is restricted to sectors 3 and 5 (since $y=0$ is
invariant). The dynamics are the same as illustrated in Figure
\ref{fig:n_a} for sectors 3 and 5, except that there is no line of
\equi points. In the case of $n\in (1,N_+)$ and $w>-1$, point
$\tilde{\mathcal{C}}_\pm$ is also physical so that we have both
matter and vacuum solutions in the state space. This is the most
interesting case and we will therefore concentrate the discussion
below to this range of $n$ for dust and radiation.

\begin{figure}[tbp]
\begin{center}
\epsfig{file=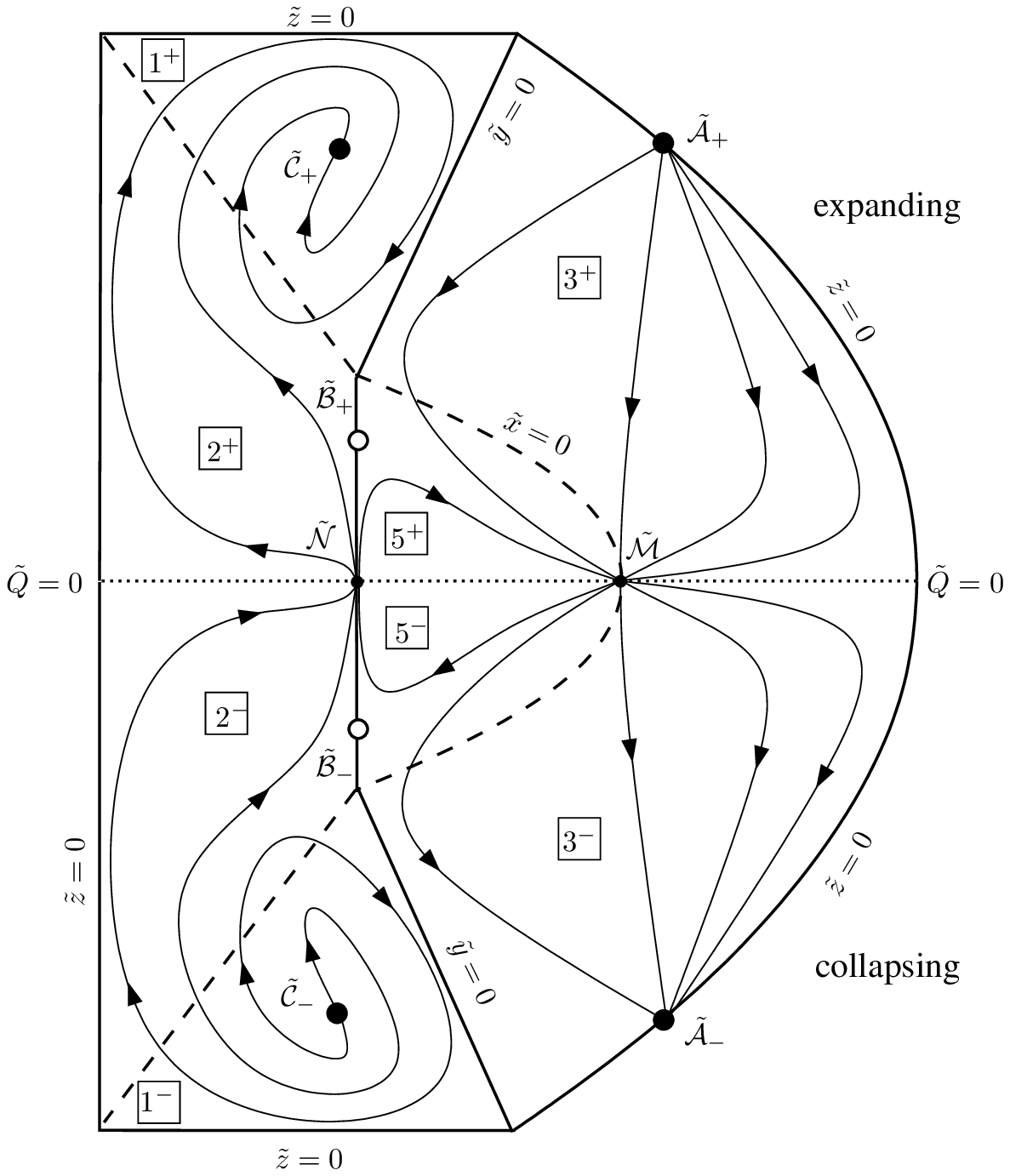, scale=0.6} \caption{Compact state space
of the flat ($k=0$) Friedmann models  with $n\in (1,N_+)$ and $w=0$.
The \equi point $\tilde{\mathcal{B}}_\pm$ is represented by a hollow
circle because it only has a corresponding cosmological solution for
the bifurcation value $w=2/3$. The dotted line represents the points
with $\tilde{Q}=0$, but only the points $\tilde{\mathcal{M}}$ with
$\tilde{x}=\tilde{Q}=0$ and $\tilde{\mathcal{N}}$ with
$\tilde{y}=\tilde{Q}=0$ may have associated solutions to the
underlying cosmological equations.}\label{fig:n_c}
\end{center}
\end{figure}

In Figure \ref{fig:n_c} we consider $w=0$ and $n\in (1,N_+)$, and it
can be seen that there are trajectories between the isotropic vacuum
points $\tilde{\mathcal{A}}_+$ and $\tilde{\mathcal{A}}_-$
corresponding to recollapsing solutions. At a first glance there
also appear to be bouncing solutions through point
$\tilde{\mathcal{N}}$ in sectors 2 and 5. In sector 2 for example,
orbits seem to move from the collapsing matter point
$\tilde{\mathcal{C}}_-$ to its expanding counterpart
$\tilde{\mathcal{C}}_+$. However, since $\tilde{\mathcal{N}}$ is a
degenerate point (see section 4.2) where $R=\Th=\dot{\Th}=0$, orbits
in the collapsing sector $2_-$ approach $\tilde{\mathcal{N}}$
asymptotically in the future while the orbits in the expanding
sector $2_+$ approach $\tilde{\mathcal{N}}$ asymptotically in the
past. These orbit cannot move through point $\tilde{\mathcal{N}}$
and do therefore not represent bounce solutions.

We note that in sector 5 bounce cosmologies exist in which we first
have expansion towards $\tilde{\mathcal{M}}$, and then asymptotic
collapse towards $\tilde{\mathcal{N}}$. As noted above, these orbits cannot cross at $\tilde{\mathcal{N}}$, otherwise re-expansion to
$\tilde{\mathcal{M}}$ with a final recollapse towards
$\tilde{\mathcal{A}}_-$ could have been possible. Thus cyclic
universes are not possible in this scenario, since it would require
passing through a degenerate state represented by the point
$\tilde{\mathcal{N}}$.

Comparing to \cite{Clifton05}, we observe that the orbits connecting
points $4$ and $6$ in \cite{Clifton05} correspond to the orbits
between points $\tilde{\mathcal{C}}_-$ and $\tilde{\mathcal{N}}$ in
the compact analysis for the case $n\in (1,N_+)$ considered here,
while the orbits connecting points $3$ and $5$ correspond to the
orbits between points $\tilde{\mathcal{C}}_+$ and
$\tilde{\mathcal{N}}$.  We observe again that the bounce and
recollapse behaviours found in \cite{Clifton05} are recovered in
this compact analysis.

The qualitative results remain unchanged when considering radiation
dominated regimes with $w=1/3$, the only difference being that point
$\tilde{\mathcal{B}}_\pm$ moves closer towards the intersection of
$\tilde{y}=0$ and $\tilde{x}=0$.

%%%%%%%%%%%%%%%%%%%%%%%%%%%%%%%%%%%%
\section{Remarks and Conclusions}
%%%%%%%%%%%%%%%%%%%%%%%%%%%%%%%%%%%%

In this work, we compared the use of compact and non-compact
variables for a dynamical systems analysis of alternative theories
of gravity. We first considered state spaces where
expansion--normalised variables were used. These
expansion--normalised variables were first introduced in the context
of flat FLRW models in GR \cite{Collins71}, where they define a
compact state space. This compactness is desirable for determining
the global behaviour of cosmological models and was one of the
original reasons for introducing expansion--normalised variables in
dynamical systems theory applied to cosmology. In \cite{Goliath99} a
method to compactify the state space of more general cosmologies was
introduced, which  was successfully applied to a modified theory of
gravity in \cite{Goheer07a}.

We here showed that when non-compact expansion normalised--variables
are used, we are restricted to expanding or contracting cosmological
models only. Static solutions and bouncing or recollapsing type
solutions lie at or approach infinity in this framework. This was
illustrated in section 3, where we compared the
expansion--normalised compact \cite{Goheer07a} and non-compact
\cite{Leach06} state spaces of LRS Bianchi I models in
$R^n$-gravity. While the works agree for the finite points in the
non--compact analysis with $y\neq 0$, discrepancies were found for
the points with $y=0$ and the points at infinity. For $y=0$, which
corresponds to the limit of vanishing Ricci curvature, differences
with respect to the existence of solutions at the given points were
observed. At infinity in the non--compact analysis \cite{Leach06},
both the occurrence of \equi points and the exact solutions at the
\equi points differs in places from the results obtained in
\cite{Goheer07a}. For example, we found that the asymptotic points
in \cite{Leach06} only have analogs in the compact analysis for
specific values of the parameter $n$. The asymptotic \equi point
$\mathcal{D}_\infty$ does not have a counterpart in \cite{Goheer07a}
at all - it only appears to be an \equi point in \cite{Leach06}
because the collapsing part of the state space is not included in
this case.

We resolved these problems and found the approach used in the
compact analysis \cite{Goheer07a} more straightforward to analyse:
Since there are no infinities in this framework, the opportunity to
miss relevant information is reduced.

In section 4 we extended our comparison of formalisms by considering
the non-compact non--expansion--normalised variables used in
\cite{Clifton05}. We compared the results of \cite{Clifton05} with
the compact analysis of \cite{Goheer07a} for the flat Friedmann
models. As in the first example, the points at infinity only have
corresponding \equi points in the compact analysis for specific
values of $n$ and/or $w$. Unlike in \cite{Leach06}, bounce and
recollapse behaviours could be investigated in the framework of
\cite{Clifton05}, and we recover these results in the compact
formalism.

We note that in both non-compact formalisms \cite{Leach06} and
\cite{Clifton05}, the \equi points at infinity are associated with a
divergence in the respective dynamical systems time variable $\tau$.
When expansion--normalised variables are used, $\tau$ diverges when
$\Th \rightarrow 0$, while in framework of \cite{Clifton05} $\tau$
diverges when the matter density becomes negligible. This divergence
however does not seem to effect the stability of the \equi points at
infinity, since the same results were found in the compact analysis.

Finally, we observe that in both non-compact analyses considered
here, there are more \equi points found than in the corresponding
compact analysis. In the non-compact analysis of \cite{Leach06} for
example, five individual \equi points and a line of points were
found for the expanding LRS Bianchi I vacuum models, while in the
corresponding compact analysis \cite{Goheer07a} only one \equi point
and a line of \equi points were found in the expanding subset.
Similarly, for the flat Friedmann model, five pairs of (expanding
and collapsing) \equi points were found in \cite{Clifton05}, while
only three pairs of points were found in the compact analysis
\cite{Goheer07a}. The detailed comparison in sections 3 and 4 of
this paper shows that there are two main reasons for the additional
\equi points in the non-compact analyses \cite{Leach06} and
\cite{Clifton05}. The first is the duplication of \equi points at
infinity: we find that in the non--compact analysis \cite{Clifton05}
two copies of same point in the finite analysis (here points $5,6$)
are created, and while static points exist only for special
parameter values in \cite{Goheer07a}, they have analogs at infinity
in \cite{Leach06} for all values of $n$ (see Tables
\ref{points:vac}, \ref{points:mat} and \ref{points:mat2}). Secondly,
some points which are classified as \equi points in the non-compact
analysis, are not \equi points in the compact analysis. In the
examples studied here, this applies to points $\mathcal{D}_\infty$
in \cite{Leach06} and $5,6$ in \cite{Clifton05}, which correspond to
$\tilde{\mathcal{M}}$ and $\tilde{\mathcal{N}}$ respectively.

Its worth noting that the analysis considered here is also
applicable to more general theories of gravity such as
$f(R)=R+\alpha R^n$. In principle, compact variables can be
constructed for such theories, but at the expense of a large number
of sectors to be analysed (as can be see from equation (18) in
\cite{Barrow06c}).

In conclusion, we have shown that it is advantageous to compactify
the state space whenever possible. The use of appropriately
constructed compact variables allows for a clear and complete
analysis including static, bouncing and recollapsing solutions and
avoids the complications caused by \equi points at infinity.

%%%%%%%%%%%%%%%%%%%
\ack
%%%%%%%%%%%%%%%%%%%

This research was supported by the National Research Foundation
(South Africa) and the Italian {\it Ministero Degli Affari Esteri-DG
per la Promozione e Cooperazione Culturale} under the joint Italy/
South Africa Science and Technology agreement. NG is funded by the
Claude Leon Foundation. We thank the referees for their useful
comments.

%%%%%%%%%%%%%%%%%%%%%%%%%%%
\section*{References}
%%%%%%%%%%%%%%%%%%%%%%%%%%%

\end{document}